\documentclass[twocolumn,showpacs,aps]{revtex4}
\usepackage{epsfig,amsmath}

\begin{document}
\title{Exact General Relativistic Thick Disks}

\author{Guillermo A. Gonz\'{a}lez}\email[e-mail: ]{guillego@uis.edu.co}

\affiliation{Escuela de F\'{\i}sica, Universidad Industrial de Santander,
A. A. 678, Bucaramanga, Colombia}

\author{Patricio S. Letelier}\email[e-mail: ]{letelier@ime.unicamp.br}

\affiliation{Departamento de Matem\'atica Aplicada, IMECC, Universidade
Estadual de Campinas, 13081-970 Campinas, S.P., Brazil}


\begin{abstract}

A method to construct exact general relativistic thick disks that is a simple 
generalization of the ``displace, cut and reflect'' method commonly used in 
Newtonian, as well as, in Einstein theory of gravitation is presented. This 
generalization consists  in  the addition of a new step in the above mentioned
method. The new method can be  pictured as a ``displace, cut, {\it fill} and
reflect'' method. In the Newtonian case, the method is  illustrated in some
detail  with the Kuzmin-Toomre  disk.  We obtain a thick disk with acceptable
physical properties. In the relativistic case two solutions of the  Weyl
equations, the Weyl gamma metric (also known as Zipoy-Voorhees metric) and the
Chazy-Curzon metric are used to construct thick disks. Also  the Schwarzschild
metric in isotropic coordinates is employed to construct another family of
thick disks. In all the considered cases  we have non trivial ranges of the
involved parameter  that yield thick disks in which all  the energy conditions
are satisfied.

\end{abstract}

\maketitle

\section{Introduction}

Exact solutions of the Einstein equations are associated  to highly idealized
physical  systems that have some exceptional geometrical properties. In some
cases with a simple exact solution  one can capture a significant part of the
physical properties of non trivial systems. Also, in nonlinear theories like
general relativity  and fluid dynamics the exact solutions play an important
role in numerical analysis. These solutions can be used to test numerical codes
and its  outcomes.  Also they can be employed  as initial conditions to
describe  more realistic situations, e.g., a static solution can be used as
part of the initial conditions  for a full dynamical simulation.

Since the natural shape of an isolated  self-gravitating fluid is axially
symmetric, the solutions of Einstein's field equations with this symmetry play
a particularly important role in the  astrophysical  applications of general
relativity. In particular disk like configurations of matter are of great 
interest, since they can be used as models of galaxies or accretion disks. Also
these disks can be used as starting point to represent  more realistic models
in which the bulge and halo of the galaxy are considered.

Solutions for static thin disks without radial pressure were first studied by
Bonnor and Sackfield \cite{BS}, and Morgan and Morgan \cite{MM1}, and with
radial pressure by Morgan and Morgan \cite{MM2}.  The first solution
represent disks made of pressureless dust whereas the second disks with
azimuthal pressure but without radial pressure. And the third a disk made of an
anisotropic fluid with nonzero radial pressure. The Bonnor-Sackfield disk has
a singular rim. These disks are finite.

Several classes of exact solutions of the Einstein field equations
corresponding to static \cite{LP,LO,LEM,LL1,BLK,BLP,LL2,LL3} and stationary
\cite{BL,LZB,GL2} thin disks have been obtained by different authors, with or
without radial pressure. Thin disks with radial tension \cite{GL1}, magnetic
fields \cite{LET1} and magnetic and electric fields \cite{KBL} have been also
studied. The non linear superposition of a disk  and a black hole was first
considered by Lemos and Letelier \cite{LL1}. This solution and its 
generalizations has been studied in some detail in
\cite{SZZ1,SZZ2,SZ1,ZS1,SZ2,SEM1,SEM2,SEM3}.  Recently the stability of
circular orbits of particles moving around black holes surrounded by axially
symmetric structures have been considered in \cite{LET2}. For a recent survey
on relativistic gravitating thin disks, see \cite{SEM4}.
       
Except for the pressureless disks all the other disks have as source matter
with azimuthal pressure (tension)  different from the radial pressure
(tension).  However, in some cases these disks can be interpreted as the
superposition of two counter-rotating perfect fluids.  A detailed study of the
counter-rotating model for the case of static thin disks  is  presented
in \cite{GAG}. Recently, more realistic models of thin disks and thin disks
with halos  made of perfect fluids were considered in \cite{DANLET}.
    
In all  the disks mentioned above an inverse style method was used to solve the
Einstein equations. The metric representing the disk is guessed and then it is
used to  compute the source  (energy-momentum tensor). This method was  named
by Synge as the g-method \cite{SYN} in contrast to the  t-method or direct
method in which  the source is given and the Einstein equations are solved.
The  t-method,  has  been used to generate disks by the Jena group
\cite{KLE1,NM,KR,KLE2,FK,KLE3,KLE4}. Essentially, they  are obtained by
solving a Riemann-Hilbert problem. These solutions are highly non trivial, but
they deserve special attention because of their clear physical meaning.

In the solutions obtained by the g-method  the well known ``displace, cut and
reflect'' method is used. The idea of the method is simple. Given a solution of
the vacuum Einstein equations, a cut is make above all singularities or
sources. The identification of this solution with its mirror image yields
relativistic models of disks. In general, these disks are of infinite extension and   finite  mass.

The aim of this paper is to consider disks beyond the thin disk limit to add a
new degree of reality to these geometric  models  of galaxies. Even though in
first  approximation the galactic disks can be considered to be  very thin , e.g., in
our Galaxy the radius of the disk is  10$kpc$ and its thickness is 1$kpc$. In a
more realistic model the thickness of the disk need to be considered. Also 
 it is well known in fluid mechanics  that the addition of a new dimension can make
dramatic  changes in the  dynamics of the fluid. In principle, 
this new dimension    will also change the
 dynamical properties  of the  the disk source, e.g., its  stability.

In  this paper we  generalize the ``displace, cut and reflect'' method in order
to obtain thick disks models from vacuum solutions of Einstein equations.  We 
shall  replace the surface of discontinuity of the metric derivatives  with a
``thick'' shell in such a way that the matter content of the disk will be
described by continuous functions  with continuous first derivatives. This 
generalization can be named ``displace, cut, {\it fill} and reflect'' method.
The disks obtained with this method, in general, will be of infinite extension
and finite mass. Also as in the case of thin  disks,                          
the matter that form the thick disks will not obey   simple equations of state and  in some regions of the disk  the pressure can change  sign given rise to  tensions. Although the energy condition will be  fulfilled. The models of thick disks presented can be considered aS generalizations of models of thin
 disks studied in references  \cite{KBL} and \cite{DANLET}.

The article is divided as follows. In Sec. II we present, in some detail, the
main idea of the ``displace, cut, {\it fill} and reflect'' method in 
 Newtonian gravity. The method is then applied, in Sec. III, to construct
relativistic thick disks in Weyl coordinates. We also study  the general
expression for the energy-momentum tensor of the disks. The method is
illustrated by taking two simple Weyl solutions that lead to thick disks in
agreement with all the energy-conditions. In Sec. IV we apply the method to the
Schwarzschild solution in isotropic cylindrical coordinates. The disk obtained
also satisfy all the energy conditions, this disk have equal azimuthal and
radial pressures and different vertical pressure. Finally, in Sec. V, we
summarize our main results.

\section{Newtonian Thick Disks}

\begin{figure*}
$$\begin{array}{cc}
\ \epsfig{width=2.25in,file=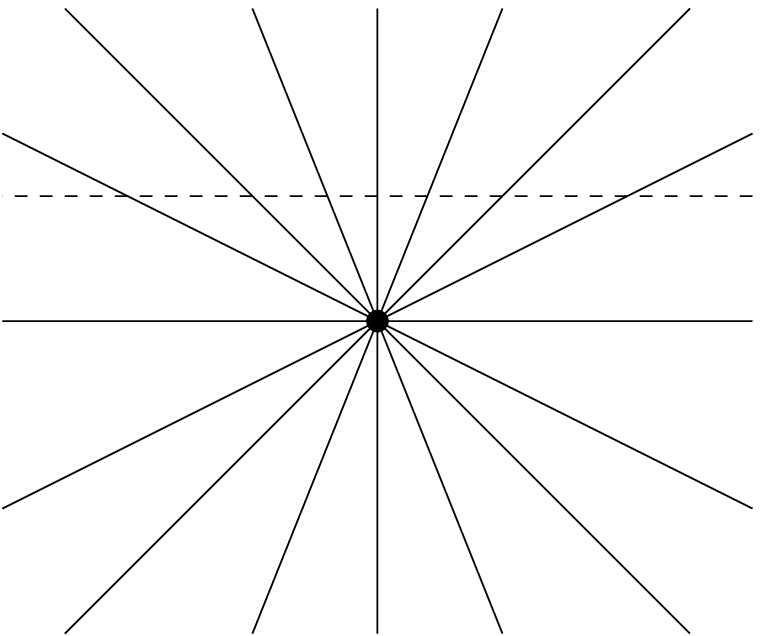} \ & 
\qquad \epsfig{width=2.25in,file=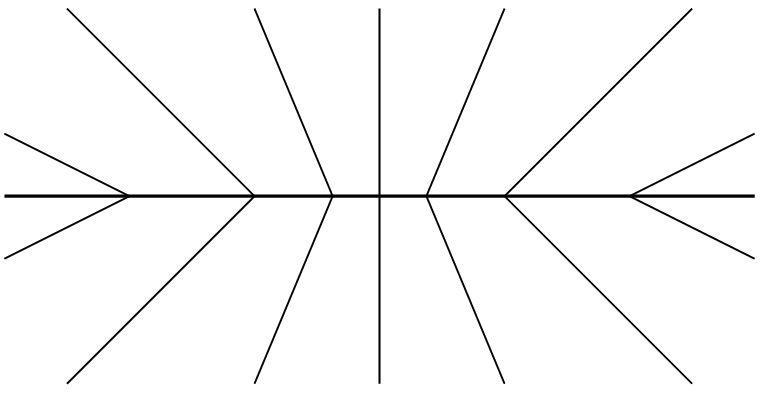} \\
(a) & (b)
\end{array}$$
\caption{Construction of a thin disk by the ``displace, cut and reflect''
method from the gravitational field of a mass point. In (a) the space with a
singularity is displaced and cut by a plane (dashed line). In (b) the part with
singularities is disregarded and the upper part is reflected on the
plane.}\label{fig:thind}
\end{figure*}

\begin{figure*}
$$\begin{array}{cc}
\ \epsfig{width=2.25in,file=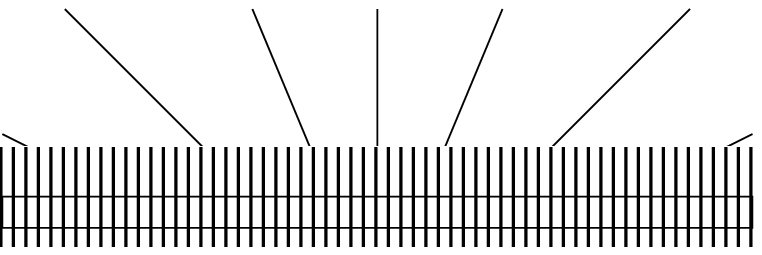} \ &
\ \epsfig{width=2.25in,file=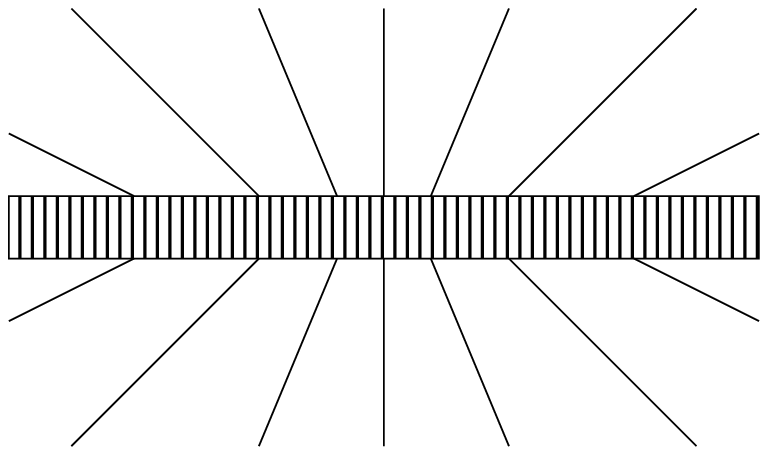} \ \\
(a) & (b)
\end{array}$$
\caption{Construction  of a thick disk by the ``displace, cut, fill and
reflect'' method. In (a), after disregard the part with singularities, we put a
thick shell below the plane. In (b), we reflect the resultant configuration on
the bottom surface of the shell.}\label{fig:thickd}
\end{figure*}

\begin{figure*}
$$\begin{array}{ccc}
\ \epsfig{width=2in,file=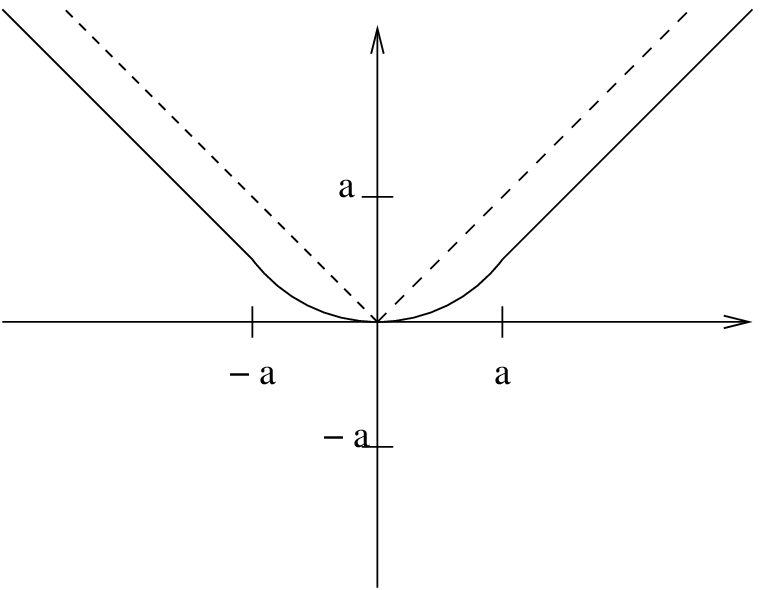} \ &
\ \epsfig{width=2in,file=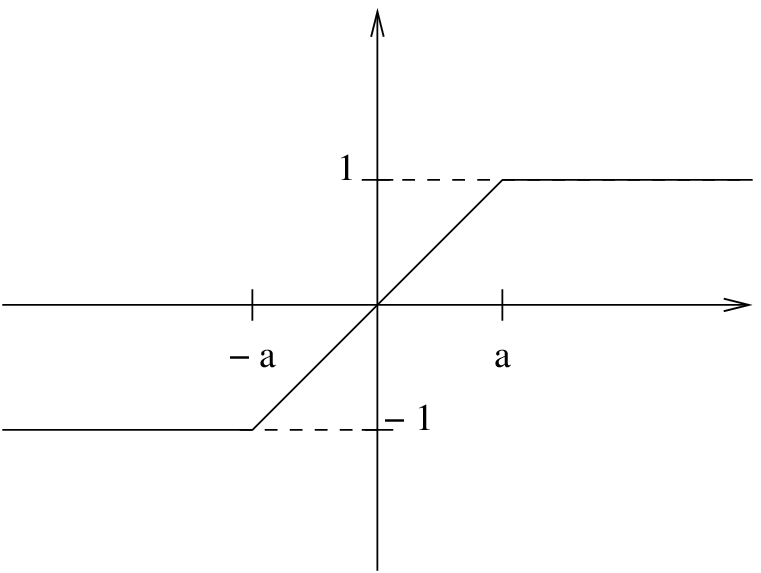} \ &
\ \epsfig{width=2in,file=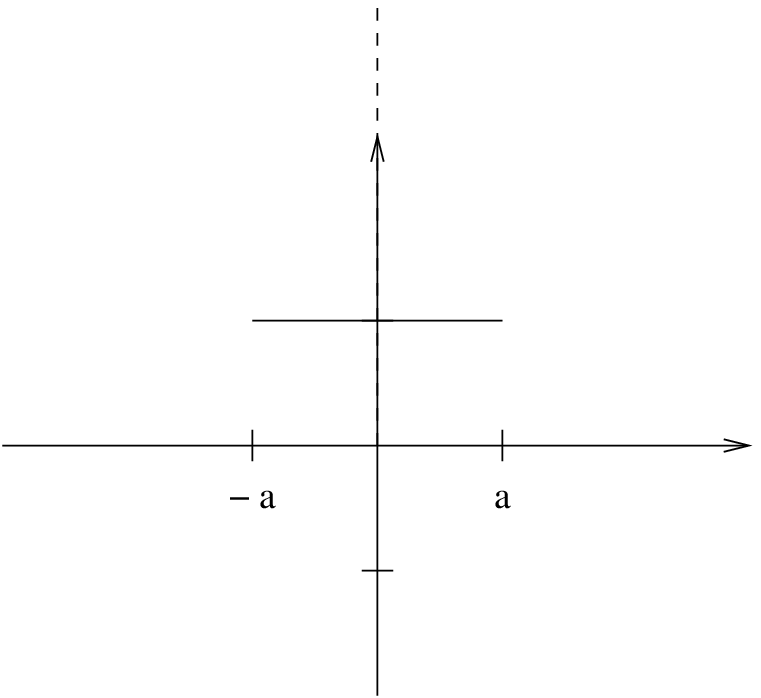} \	\\
(a) & (b) & (c)
\end{array}$$
\caption{The function $h(z)$ and its derivatives, for thin disks (dashed line)
and thick disks (solid line). In (a) we plot $h (z)$, in (b) $h' (z)$ and in
(c) $h'' (z)$.}\label{fig:figh}
\end{figure*}

The Newtonian gravitational potential of a thin disk can be obtained by a
simple procedure, the ``displace, cut and reflect method'', due to Kuzmin
\cite{KUZ}.  The method can be divided in the following steps: First, choose a
surface that divides the usual space in two parts: one with no singularities or
sources an the other with the sources. Second, disregard the part of the space
with singularities. Third, use the surface to make an inversion of the
nonsingular part of the space. The result will be a space with a singularity
that is a delta function with support on $z = 0$. This procedure is depicted in
Fig. \ref{fig:thind}.

In order to obtain a thick disk we need to modify the above procedure.  We
essentially need  to replace the surface of discontinuity with a ``thick''
shell in such a way that the matter content of the disk  be described by
continuous functions. Now  method has an additional step and  
 can be named ``displace, cut, {\it fill}
and reflect''. After we disregard the part of the
space with singularities, we put a thick shell below the surface. Then we use
the bottom surface of the shell to make the inversion. The procedure is
illustrated in Fig. \ref{fig:thickd}.

Mathematically the method is equivalent to make the transformation $z \to h (z)
+ b$,  where $b$ is a  constant and $h (z)$ an even function of
$z$. In Newtonian gravity, the potential $\Phi (r,z)$ is a solution of the
Laplace Equation 
\begin{equation}
\nabla^2 \Phi = \Phi_{,rr} + \frac{\Phi_{,r}}{r} + \Phi_{,zz} = 0 ,
\end{equation}
where $(r,\varphi,z)$ are the usual cylindrical coordinates. After we
make the transformation $z \to h (z) + b$, the above equation leads to
\begin{equation}
\nabla^2 \Phi = h'' \Phi_{,h} + [(h')^2 - 1] \Phi_{,hh} , \label{eq:laph}
\end{equation}
where primes indicate differentiation with respect to $z$.

For the case of thin disks we take $h (z) = |z|$. Note that $\partial_z |z| =
2 \theta (z) - 1$ and $\partial_z \theta (z) = \delta (z)$ where $\theta
(z)$  is the Heaveside function and $\delta (z)$ the usual Dirac distribution.
By using (\ref{eq:laph}), the Poisson equation leads to a mass density given by
\begin{equation}
2 \pi G \rho (r,z) = \Phi_{,h} \delta (z).
\end{equation}
We have a surface distribution of matter located in the plane $z = 0$. In Fig
\ref{fig:figh}, with dashed lines, we plot $h (z)$ an its derivatives.

For the case of thick disks the function $h(z)$ must be selected in such a way
that $\Phi$ and its first derivatives  be continuous across the plane $z =
0$. Let us take a function $h (z)$ defined as
\begin{equation}
h (z) = \left\{
	\begin{array}{rcr}
	z - a/2 & , & z \geq a ,\\
		& &	\\
	z^2/2a & , & - a \leq z \leq a ,\\
		& &	\\
	- z - a/2 & , & z \leq - a .
	\end{array} \right . \label{eq:funh}
\end{equation}
Hence, by taking the function $h(z)$ above defined we can generate disks of
thickness $2a$ located in the region $-a \leq z \leq a$. In Fig.
\ref{fig:figh}  we plot $h (z)$ an its derivatives (solid lines).

When  $|z| \geq a$ the function $h (z)$ is a linear function of $z$ such that
  $h' (z) = 1$. Hence, its  second derivative is  zero. Then  the mass density
vanish outside the disk. Since the first derivative is continuous at $|z| = a$
(see Fig. \ref{fig:figh}) and  the second derivative  piecewise constant we
have that  the mass density, $\rho$, will be well defined inside the disk,
\begin{equation}
4 \pi G a^2 \rho (r,z) = a \Phi_{,h} + (z^2 - a^2) \Phi_{,hh} \ ,
\end{equation}
for $|z| \leq a$ and $\rho=0$ for $|z| > a$.

As a simple example we can consider the usual potential for a mass point,
written in cylindrical coordinates as
\begin{equation}
\Phi = - \frac{Gm}{\sqrt{r^2 + z^2}} .
\end{equation}
By doing  the transformation $z \to h (z) + b$ in the previous potential,
 we obtain the mass
densities
\begin{equation}
\rho (r,z) = \frac{ m \ b \ \delta (z) }{2 \pi ( r^2 + b^2 )^{3/2}} ,
\label{thinnden}
\end{equation}
for a thin disk (Kuzmin-Toomre disk \cite{KUZ,TOO}), and
\begin{eqnarray}
\rho (r,z) &=& \frac{m [3z^2 + 2a(b - a)]}{8 \pi a^2 {\rm R}^3} 
\nonumber \\
&&	\label{thicknden}\\
&+& \frac{3m(a^2 - z^2)(z^2 + 2ab)^2}{16 \pi a^4 {\rm R}^5} , \nonumber 
\end{eqnarray}
where ${\rm R}^2 = r^2 + (h + b)^2$, for a  thick disk. When $b \geq a$,
the mass density will be positive everywhere.

The function $h(z)$ presented in (\ref{eq:funh}) is the simplest function that
have the desired properties. The part of the function in the domain $|z| \leq
a$ can be changed by superposition of even functions of $z$. In  $|z| = a$  
this new function need to be matched  continuously with linear functions of
$z$ such that $h' (z) = 1$, like in  (\ref{eq:funh}). In this case the mass
density can depend on the variable $z$ in  a more general way.

It is instructive to obtain the surface
thin  disk density (say  $\sigma$ ) associated to (\ref{thinnden}),
\begin{equation}
 \sigma =\frac{ m \ b  }{2 \pi ( r^2 + b^2 )^{3/2}},\label{sigma}
\end{equation}
as a  limit of the thick disk volume density (\ref{thicknden}). To obtain the surface density of the thin disk  from  the volume density
(\ref{thinnden}) we first do 
\begin{equation}
\Sigma= 2a \rho(r,z). \label{Sig} 
\end{equation}
Now to perform the thin disk limit we put  $z=\alpha \xi$ and $a=\beta \xi$ 
($\alpha$ and $\beta$ artitrary constants) in $\Sigma$ and  we take the limit
$
\lim_{\xi \rightarrow 0 }\Sigma.
$
 This limit is just (\ref{sigma}).

\section{Relativistic Thick Disks in Weyl Coordinates}

When the matter is absent, the metric for  a static axially symmetric
 spacetime  can be cast  without loosing  generality as
\begin{equation}
ds^2 = - e^{2\Phi} dt^2 + e^{- 2\Phi} [ r^2 d\varphi^2 + e^{2\Lambda} ( dr^2 +
dz^2 ) ] , \label{eq:metw}
\end{equation} 
where $\Phi$ and $\Lambda$ are functions of $r$ and $z$ only. The ranges of the
coordinates $(\varphi,r,z)$ are the usual for cylindrical coordinates (Weyl
coordinates) and $-\infty  \leq t<\infty$.  The Einstein vacuum equations for this
metric yield the Weyl equations \cite{WEY1,WEY2},
\begin{subequations}\begin{eqnarray}
&&\Phi_{,rr} + \frac{\Phi_{,r}}{r} + \Phi_{,zz} = 0 , \\
&&\nonumber	\\
&&\Lambda_{,r} = r ( \Phi_{,r}^2 - \Phi_{,z}^2 ) , \\
&&\nonumber	\\
&&\Lambda_{,z} = 2 r \Phi_{,r} \Phi_{,z} .
\end{eqnarray}\label{eq:eqsw}\end{subequations}

From a  solution of Einstein vacuum equations corresponding to a Weyl
metric (\ref{eq:metw}), we can construct a thick disk model by means of the
``displace, cut, fill and reflect method'' using the transformation $z \to h
(z) + b$, with $h (z)$ given by (\ref{eq:funh}). The energy-momentum tensor of
the disk can be computed using the Einstein equations in the matter, written as
\begin{equation}
T_{ab} = R_{ab} - \frac{1}{2} g_{ab} R , \label{eq:eqm}
\end{equation}
in units such that $c = 8\pi G = 1$.

By using the Einstein equations (\ref{eq:eqsw}), the nonzero components of
$T_a^b$ are:
\begin{subequations}\begin{eqnarray}
T_t^t &=& \frac{e^{2(\Phi - \Lambda)}}{a^2} [ a (\Lambda_{,h} - 2 \Phi_{,h}) +
\nonumber \\
& & 	\\
& &( z^2 - a^2 ) (\Phi_{,h}^2 - 2 \Phi_{,hh} + \Lambda_{,hh} ) ] , \nonumber \\
\nonumber \\
T_\varphi^\varphi &=& \frac{e^{2(\Phi - \Lambda)}}{a^2} [ a \Lambda_{,h} + (
z^2 - a^2 ) ( \Phi_{,h}^2 + \Lambda_{,hh} ) ] , \\
	\nonumber	\\
T_r^r &=& \frac{e^{2(\Phi - \Lambda)}}{a^2} ( z^2 - a^2 ) \Phi_{,h}^2 \ , \\
	\nonumber	\\
T_z^z &=& \frac{e^{2(\Phi - \Lambda)}}{a^2} ( a^2 - z^2 ) \Phi_{,h}^2 \ ,
\end{eqnarray}\label{eq:emtw}\end{subequations}
valid for the region $- a \leq z \leq a$. Outside this region we have $T_a^b =
0$.

For the above expressions we can see that the radial stress $T_r^r$ is
negative (we have radial tension). On the other hand, since
 $T_z^z = - T_r^r$, we
have vertical pressure. Defining the  orthonormal tetrad $\{
V^b , X^b , Y^b , Z^b \}$, where
\begin{subequations}\begin{eqnarray}
V^a &=& e^{- \Phi} \ ( 1, 0, 0, 0 ) \ ,	\\
	&	&	\nonumber	\\
X^a &=& \frac{e^\Phi}{r} \ \ ( 0, 1, 0, 0 ) \ ,	\\
	&	&	\nonumber	\\
Y^a &=& e^{\Phi - \Lambda} ( 0, 0, 1, 0 ) \ ,	\\
	&	&	\nonumber	\\
Z^a &=& e^{\Phi - \Lambda} ( 0, 0, 0, 1 ) \ ,
\end{eqnarray}\label{eq:tetrad}\end{subequations}
we can be cast  the energy-momentum tensor in its canonical form
\begin{equation}
T_{ab} = \epsilon V_a V_b + p_\varphi X_a X_b  + p_r Y_a Y_b + p_z Z_a Z_b \ .
\label{eq:canw}
\end{equation}
Here $\epsilon = - T_t^t$ is the energy density, $p_\varphi =
T_\varphi^\varphi$ is the azimuthal stress, $p_r = T_r^r = - T_z^z$ is the
radial tension and $p_z = T_z^z$ is the vertical pressure.

From (\ref{eq:emtw}) we get 
 the ``effective Newtonian'' density, $\rho = \epsilon + p_\varphi + p_r +
p_z = \epsilon + p_\varphi,$  
\begin{equation}
\rho = \frac{2 e^{2(\phi - \Lambda)}}{a^2} [ a \Phi_{,h} + (z^2 - a^2)
\Phi_{,hh} ]. \label{eq:dnew}
\end{equation}
 The strong energy condition requires that $\rho
\geq 0$, whereas the weak energy condition imposes the condition $\epsilon \geq
0$. The dominant energy condition is equivalent to the requirement
$|\frac{p_\varphi}{\epsilon}| \leq 1$, $|\frac{p_r}{\epsilon}| \leq 1$ and
$|\frac{p_z}{\epsilon}| \leq 1$, see for instance \cite{HE}.

One can obtain the quantities associated to a general relativistic 
thin disk from the
 corresponding 	quantities associated to the thick disk using the same limit
 procedure described in the Newtonian disk  case.

\subsection{Thick Disks from the Chazy-Curzon Metric}

As a first example we apply the ``displace, cut, fill and reflect'' method to
obtain thick disks using the Chazy-Curzon solution \cite{CHA,CUR}, written in
Weyl coordinates as
\begin{subequations}\begin{eqnarray}
\Phi &=& - \frac{m}{\rm R} ,	\\
&&	\nonumber	\\
\Lambda &=& - \frac{m^2 r^2}{2 {\rm R}^4} ,
\end{eqnarray}\end{subequations}
where ${\rm R}^2 = r^2 + (h + b)^2$, $m$ and $b$ are positive constants and $h
(z)$ is given by (\ref{eq:funh}).

 Now  we  rescale the variables and the parameters in terms of the disk 
thickness,
$a$. We  $r = a \tilde r$, $z = a \tilde z$, ${\rm R} = a
\tilde{\rm R}$, $b = a \tilde b$ and $m = a \tilde m$.  From
(\ref{eq:emtw}) and (\ref{eq:dnew}), we obtain
\begin{widetext}
\begin{subequations}\begin{eqnarray}
\tilde\rho &=& \frac{ \tilde m \ e^{2(\Phi - \Lambda)}}{2 \tilde{\rm R}^5}
\left[ 2 ( 3 {\tilde z}^2 + 2 (\tilde b - 1) ) \tilde{\rm R}^2 + 3 (1 - {\tilde
z}^2) ({\tilde z}^2 + 2 \tilde b)^2 \right] , \label{eq:dncc} \\
&&	\nonumber	\\
\tilde\epsilon &=& \frac{\tilde m \ e^{2(\Phi - \Lambda)}}{4 \tilde{\rm R}^8}
\left[ 4 ( 3 {\tilde z}^2 + 2 (\tilde b - 1) ) ( \tilde{\rm R}^3 - \tilde m
{\tilde r}^2 ) \tilde{\rm R}^2 + [ 6 ( \tilde{\rm R}^3 - 2 \tilde m {\tilde
r}^2 ) + \tilde m \tilde{\rm R}^2 ) (1 - {\tilde z}^2) ({\tilde z}^2 + 2 \tilde
b)^2 \right] , \label{eq:encc} \\
{\tilde p}_\varphi &=& \frac{{\tilde m}^2 e^{2(\Phi - \Lambda)}}{4 \tilde{\rm
R}^8} \left[ 4 {\tilde r}^2 [ ( 3 {\tilde z}^2 + 2 (\tilde b - 1) ) \tilde{\rm
R}^2 + 3 (1 - {\tilde z}^2) ({\tilde z}^2 + 2 \tilde b)^2 ] + ({\tilde z}^2 -
1) ({\tilde z}^2 + 2 \tilde b)^2 \tilde{\rm R}^2 \right] , \label{eq:p1cc} \\
&&	\nonumber	\\
{\tilde p}_r &=& \frac{{\tilde m}^2 e^{2(\Phi - \Lambda)}}{4 \tilde{\rm R}^6}
\left[ ({\tilde z}^2 - 1) ({\tilde z}^2 + 2 \tilde b)^2 \right] , \label{p2cc}
\\
&&	\nonumber	\\
{\tilde p}_z &=& \frac{{\tilde m}^2 e^{2(\Phi - \Lambda)}}{4 \tilde{\rm R}^6}
\left[ (1 - {\tilde z}^2) ({\tilde z}^2 + 2 \tilde b)^2 \right] , \label{p3cc}
\end{eqnarray}\end{subequations}
\end{widetext}
where $\tilde\rho = a^2\rho$, $\tilde\epsilon = a^2\epsilon$ and ${\tilde
p}_i = a^2 p_i$.

Equation (\ref{eq:dncc}) and the condition $\tilde b \geq 1$ imply  $\rho >
0$.  Then, when  $\tilde b \geq 1$ the strong energy condition is satisfied. 
From (\ref{eq:encc}) we conclude that $\epsilon \geq 0$ whenever $\tilde {\rm
R}^3 \geq 2 \tilde m {\tilde r}^2$ and $\tilde b \geq 1$.  From the definition
of $R$ we conclude that  in order to have $\epsilon \geq 0$ everywhere, we need
$$
0 \leq \tilde m \leq \frac{\sqrt{3} \tilde b}{2} ,
$$
and $\tilde b \geq 1$.

\begin{figure*}
$$\begin{array}{cc}
\ \epsfig{width=2.75in,file=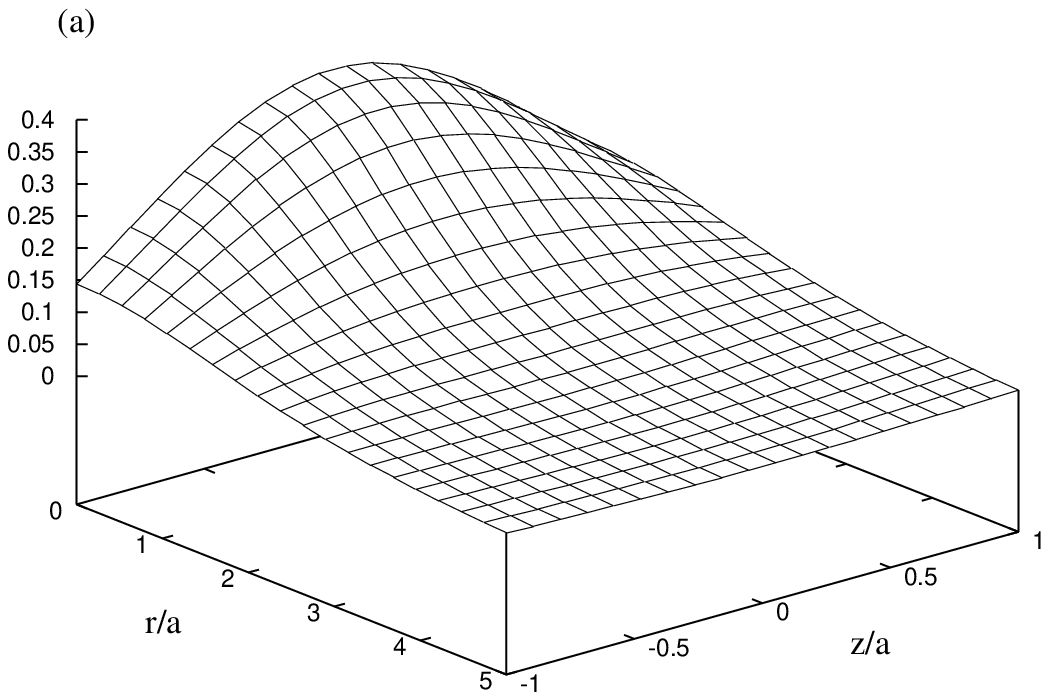} \ &
\ \epsfig{width=2.75in,file=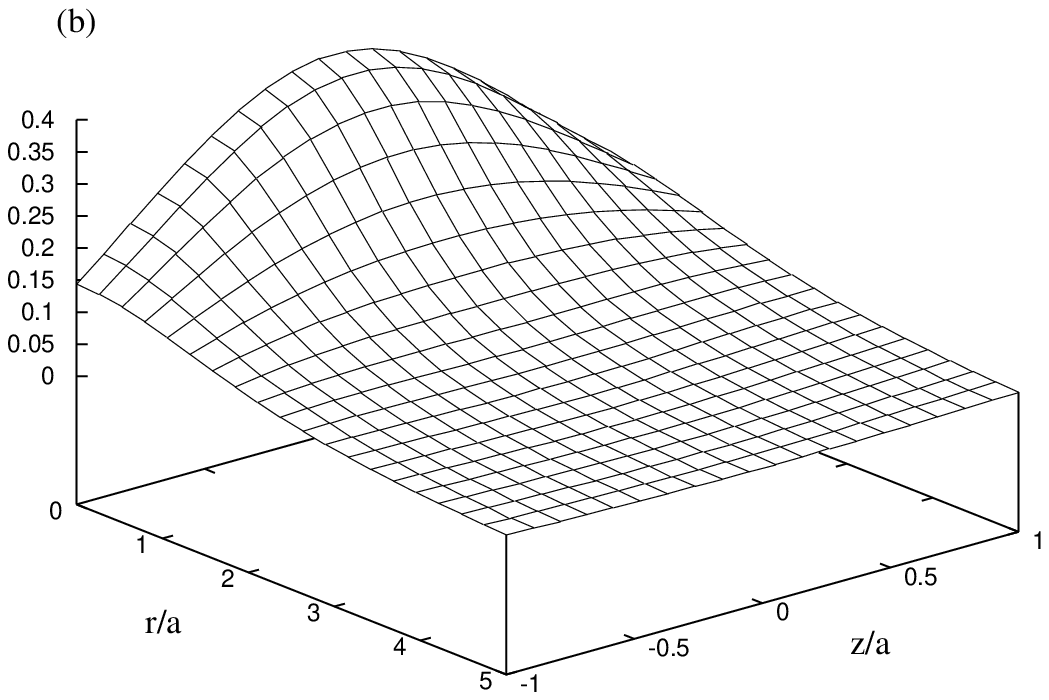} \
\end{array}$$
\caption{For a thick disk obtained from the Chazy-Curzon solution with $\tilde
m = 1$ and $\tilde b = 2$ we plot (a) the dimensionless Newtonian density
$\tilde\rho$ and (b) the dimensionless energy density $\tilde\epsilon$, as
functions of $\tilde r$ and $\tilde z$.}\label{fig:dens1}
\end{figure*}

\begin{figure*}
$$\begin{array}{cc}
\ \epsfig{width=2.75in,file=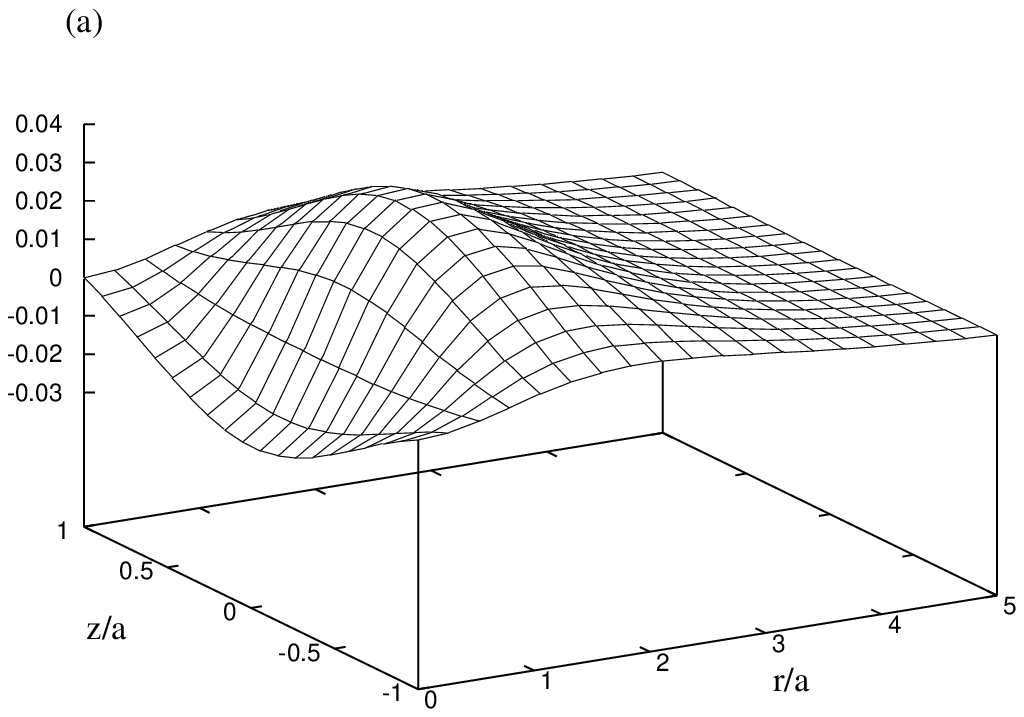} \ &
\ \epsfig{width=2.75in,file=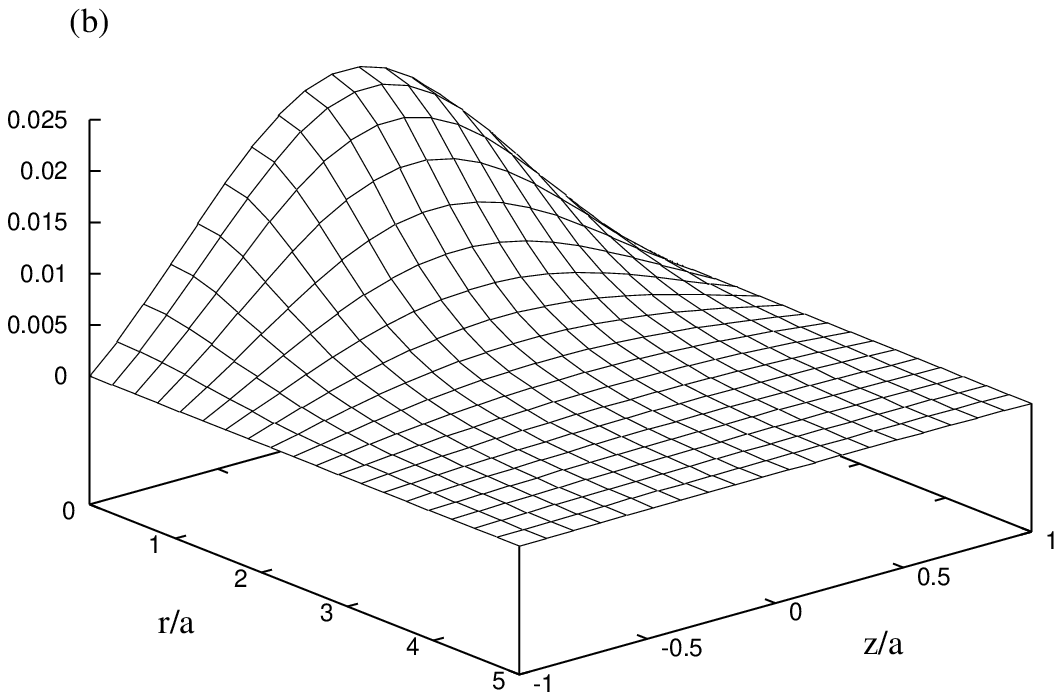} \ 
\end{array}$$
\caption{For a thick disk obtained from the Chazy-Curzon solution with $\tilde
m = 1$ and $\tilde b = 2$ we plot (a) the dimensionless azimuthal pressure
${\tilde p}_\varphi$ and (b) the dimensionless vertical pressure ${\tilde
p}_z$.}\label{fig:pres1}
\end{figure*}

The behavior of the densities is better illustrated graphically. In order to
have a disk in agreement with the weak and strong energy conditions, we take
$\tilde m = 1$ and $\tilde b = 2$. In Fig. \ref{fig:dens1} we plot the
effective Newtonian density and the energy density in units of $a^2$, $\tilde
\rho$ and $\tilde \epsilon$, as functions of $\tilde r$ and $\tilde z$. We can
see that the densities have a maximum at the center of the $z = 0$ plane. Then
the densities decrease monotonously as $r$ increase and also the densities
decrease for $z \to \pm a$. We  see that $\rho$ and $\epsilon$ have similar
magnitudes.

The azimuthal stress, as we can see from (\ref{eq:p1cc}), is negative at the
center of the disk ($r = 0$), whereas it is positive for larges values of $r$.
The boundary between the region of negative stress and positive stress is the
surface
\begin{equation}
f (r,z) = \frac{4 {\rm R}^8 e^{2(\Lambda - \Phi)}}{m^2 a^2} {\tilde p}_\varphi
= 0 \ ,
\label{eq:surf}
\end{equation}
with ${\tilde p}_\varphi$ given by (\ref{eq:p1cc}).

The behavior of the stresses is also better illustrated graphically. In  Fig.
\ref{fig:pres1} we plot the  dimensionless  azimuthal stress ${\tilde
p}_\varphi$ and vertical pressure ${\tilde p}_z$, as functions of $\tilde r$
and $\tilde z$.  The radial pressure is given by $p_r = - p_z$. Again we
take $\tilde m = 1$ and $\tilde b = 2$. The azimuthal stress is negative at the
central region of the disk, then increases to have a positive maximum on  the 
$z=0$  plane, for a value of $r \approx 1.5 a$. Finally, it 
 decreases monotonously
for increasing $r$ and also for $z \to \pm a$. The behavior of the vertical pressure
is like of the densities, with a maximum at the center of the $z=0$ plane and
then monotonously decreasing for increasing $r$. Also, $p_z = 0$ for $z = \pm
a$. From Figs. \ref{fig:dens1} and \ref{fig:pres1} we can also see that the
magnitude of the stresses is about a tenth of the magnitude of the densities.
We have
\begin{eqnarray*}
\left| \frac{p_\varphi}{\epsilon} \right| &\leq& 0.1 , \\
	\\
\left| \frac{p_r}{\epsilon} \right| = \left| \frac{p_z}{\epsilon} \right|
&\leq& 0.1 ,
\end{eqnarray*}
and so the disks are also in agreement with the dominant energy condition. Thin
disks based on the Chazy-Curzon metric were studied in Ref. \cite{BLK}.

\subsection{Thick Disks from the Zipoy-Voorhees Metric}

As a second example we take the Weyl gamma metric also known as Zipoy-Voorhees
solution \cite{ZIP,VOO}, that in Weyl  coordinates can be cast as \cite{RN}
\begin{subequations}\begin{eqnarray}
\Phi &=& \frac{m}{2 k} \ln \left[ \frac{{\rm R}_1 + {\rm R}_2 - 2
k}{{\rm R}_1 + {\rm R}_2 + 2 k} \right] ,\\
&&	\nonumber	\\
\Lambda &=& \frac{m^2}{2 k^2} \ln \left[ \frac{({\rm R}_1 + {\rm R}_2)^2
- 4 k^2}{4 {\rm R}_1 {\rm R}_2} \right],
\end{eqnarray}\end{subequations}
where  ${\rm R}^2_1 = r^2 + (h + b + k)^2$, ${\rm R}^2_2 = r^2 + (h + b -
k)^2$, $m$, $b$ and $k$ are positive constants and $h (z)$ is given
by (\ref{eq:funh}). When $k = m$ this solution leads to the Schwarzschild
metric and when $k \to 0$ to the Chazy-Curzon solution of the previous
section.

By using (\ref{eq:emtw}) and (\ref{eq:dnew}) we obtain, in terms of the
dimensionless variables used in the previous section,
\begin{widetext}\begin{subequations}\begin{eqnarray}
\tilde\rho &=& \frac{\tilde m \ e^{2(\Phi - \Lambda)}}{2 \tilde k \tilde{\rm
R}^3_1 \tilde{\rm R}^3_2} \left\{ 2 (\tilde{\rm R}_1 - \tilde{\rm R}_2)
\tilde{\rm R}^2_1 \tilde{\rm R}^2_2 + (1 - {\tilde z}^2) \left[ ({\tilde z}^2 +
2 \tilde b) (\tilde{\rm R}^3_1 - \tilde{\rm R}^3_2) - 2 \tilde k (\tilde{\rm
R}^3_1 + \tilde{\rm R}^3_2)\right] \right\}, \label{eq:dnzv} \\
&&	\nonumber	\\
\tilde\epsilon &=& \tilde\rho \left[ 1 - \frac{2 \tilde m {\tilde r}^2
(\tilde{\rm R}_1 + \tilde{\rm R}_2)}{\tilde{\rm R}^2_1 \tilde{\rm R}^2_2
[(\tilde{\rm R}_1 + \tilde{\rm R}_2)^2 - 4 {\tilde k}^2]} \right] +
\frac{{\tilde m}^2 e^{2(\Phi - \Lambda)}(1 - {\tilde z}^2)(\tilde{\rm R}_1 -
\tilde{\rm R}_2)}{4 {\tilde k}^2 \tilde{\rm R}^4_1 \tilde{\rm R}^4_2} 
\left[(\tilde{\rm R}_1 - \tilde{\rm R}_2) \tilde{\rm R}^2_1 \tilde{\rm R}^2_2 -
2 {\tilde r}^2 (\tilde{\rm R}^3_1 - \tilde{\rm R}^3_2) \right], \label{eq:enzv}
\\
&&	\nonumber	\\
{\tilde p}_\varphi &=& \frac{2 \tilde m \tilde\rho {\tilde r}^2 (\tilde{\rm
R}_1 + \tilde{\rm R}_2)}{\tilde{\rm R}^2_1 \tilde{\rm R}^2_2 [(\tilde{\rm R}_1
+ \tilde{\rm R}_2)^2 - 4 {\tilde k}^2]} + \frac{{\tilde m}^2 e^{2(\Phi -
\Lambda)}({\tilde z}^2 - 1)(\tilde{\rm R}_1 - \tilde{\rm R}_2)}{4 {\tilde k}^2
\tilde{\rm R}^4_1 \tilde{\rm R}^4_2}  \left[(\tilde{\rm R}_1 - \tilde{\rm R}_2)
\tilde{\rm R}^2_1 \tilde{\rm R}^2_2 - 2 {\tilde r}^2 (\tilde{\rm R}^3_1 -
\tilde{\rm R}^3_2) \right], \label{eq:p1zv} \\
&&	\nonumber	\\
{\tilde p}_r &=& \frac{{\tilde m}^2 e^{2(\Phi - \Lambda)}}{\tilde{\rm R}^2_1
\tilde{\rm R}^2_2} \left[ \frac{({\tilde z}^2 - 1) ({\tilde z}^2 + 2 \tilde
b)^2}{(\tilde{\rm R}_1 + \tilde{\rm R}_2)^2} \right] , \\ \label{p2zv}
&&	\nonumber	\\
{\tilde p}_z &=& \frac{{\tilde m}^2 e^{2(\Phi - \Lambda)}}{\tilde{\rm R}^2_1
\tilde{\rm R}^2_2} \left[ \frac{(1 - {\tilde z}^2) ({\tilde z}^2 + 2 \tilde
b)^2}{(\tilde {\rm R}_1 + \tilde {\rm R}_2)^2} \right] , \label{p3zv}
\end{eqnarray}\end{subequations}
\end{widetext}
where $k = a \tilde k$, ${\rm R}_1 = a \tilde{\rm R}_1$ and ${\rm R}_2 = a
\tilde{\rm R}_2$.

From (\ref{eq:dnzv}) and the  condition
$$
A \equiv \frac{\tilde k (\tilde{\rm R}^3_1 + \tilde{\rm R}^3_2)}{(\tilde{\rm
R}_1 - \tilde{\rm R}_2) \tilde{\rm R}^2_1 \tilde{\rm R}^2_2} \leq 1 ,
$$
we have  $\rho \geq 0$, i.e., the  strong energy condition is satisfied. Also
from  $\tilde{\rm R}_1^2 - \tilde{\rm R}_2^2 = 4 \tilde k (\tilde h + \tilde
b)$, with $\tilde h (\tilde z) = {\tilde z}^2 /2$, we can show that
\begin{eqnarray*}
A &=& \frac{(\tilde{\rm R}_1 + \tilde{\rm R}_2)(\tilde{\rm R}_1^3 + \tilde{\rm
R}_2^3)}{4 (\tilde h + \tilde b)\tilde{\rm R}_1^2 \tilde{\rm R}_2^2} \\ &<&
\frac{\tilde{\rm R}_1^2}{(\tilde h + \tilde b)\tilde{\rm R}_2^2} =
\frac{\tilde{\rm R}_2^2 + 4 \tilde k (\tilde h + \tilde b)}{(\tilde h + \tilde
b)\tilde{\rm R}_2^2}.
\end{eqnarray*}
Also  $\tilde h + \tilde b \geq \tilde b$ and $\tilde{\rm R}_2 \geq \tilde b -
\tilde k$ give us
$$
A < \frac{ (\tilde b - \tilde k)^2 + 4 \tilde k \tilde b}{\tilde b (\tilde b -
\tilde k)^2} .
$$
Then the condition $A \leq 1$ can be cast as
\begin{equation}
( \tilde b + \tilde k )^2 < \tilde b ( \tilde b - \tilde k )^2 ,
\label{eq:cond1}
\end{equation}
that leads also to $\tilde b \neq \tilde k$. This last condition assure the
nonsingular behavior of the energy density $\epsilon$ and the azimuthal stress
$p_\varphi$.

From (\ref{eq:enzv}) and  $r = 0$ we have that  $\epsilon > 0$. When
$\tilde z = 1$  the condition
$$
B \equiv \frac{2 \tilde m {\tilde r}^2 (\tilde{\rm R}_1 + \tilde{\rm
R}_2)}{\tilde{\rm R}^2_1 \tilde{\rm R}^2_2 [(\tilde{\rm R}_1 + \tilde{\rm
R}_2)^2 - 4 {\tilde k}^2]} \leq 1 ,
$$
gives us   $\epsilon > 0$ . Since  $\tilde {\rm R}_1 > \tilde{\rm R}_2$,
$\tilde{\rm R}_1 + \tilde {\rm R}_2 \geq 2 \tilde b$, $\tilde {\rm R}_2 >
\tilde r$ and $\tilde {\rm R}_1 \geq \tilde b + \tilde k$ we have,
$$
B < \frac{\tilde m}{({\tilde b}^2 - {\tilde k}^2) (\tilde b + \tilde k)} .
$$
Then the condition $B \leq 1$ leads to
\begin{equation}
0 < \tilde m < ({\tilde b}^2 - {\tilde k}^2) (\tilde b + \tilde k) .
\label{eq:cond2}
\end{equation}
This relation yields also $\tilde b > \tilde k$ as a condition to have $m > 0$.

\begin{figure*}
$$\begin{array}{cc}
\ \epsfig{width=2.75in,file=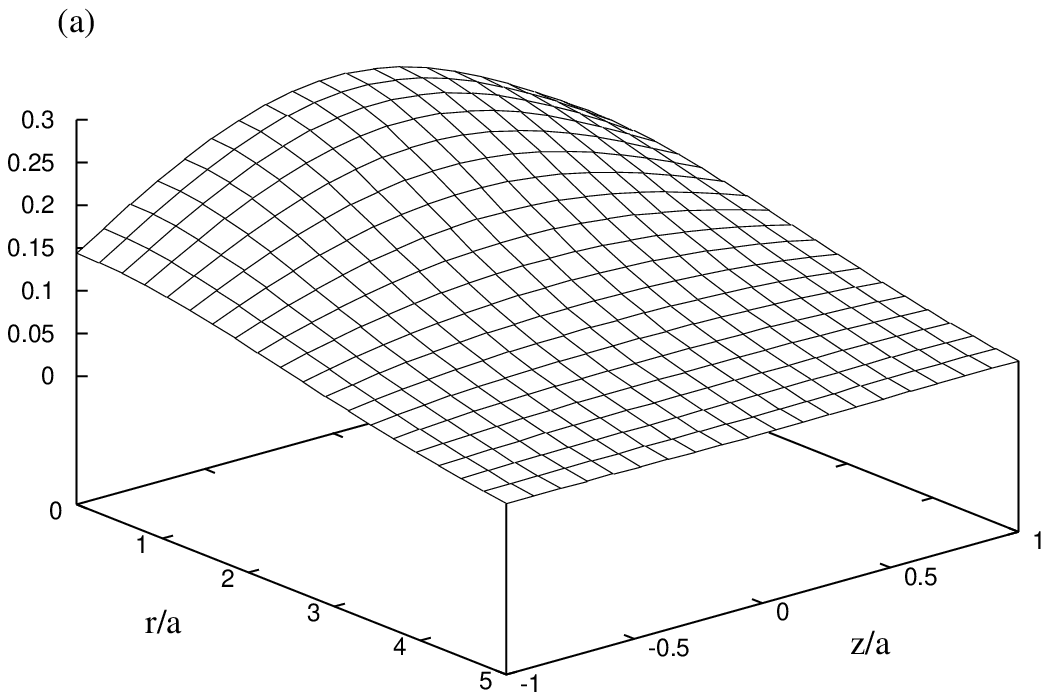} \ &
\ \epsfig{width=2.75in,file=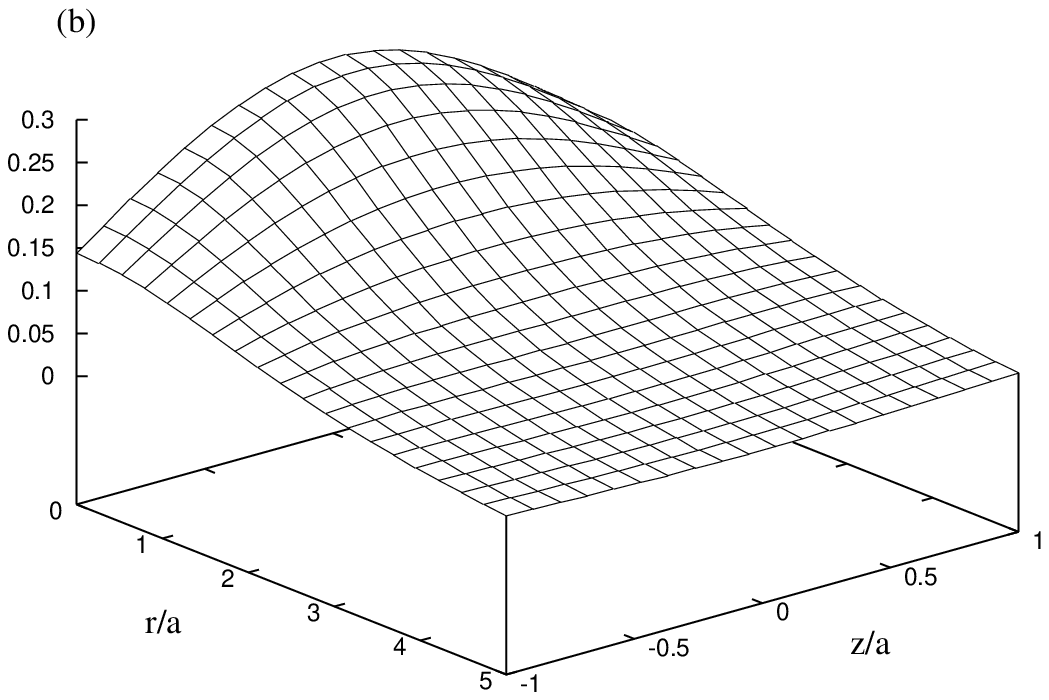} \ 
\end{array}$$
\caption{For a thick disk obtained from the Zipoy-Voorhees solution with
$\tilde m = 3$, $\tilde b = 3.5$ and $\tilde k = 1$ we plot (a) the effective
Newtonian density $\tilde\rho$ and (b) the energy density
$\tilde\epsilon$.}\label{fig:dens2}
\end{figure*}

\begin{figure*}
$$\begin{array}{cc}
\ \epsfig{width=2.75in,file=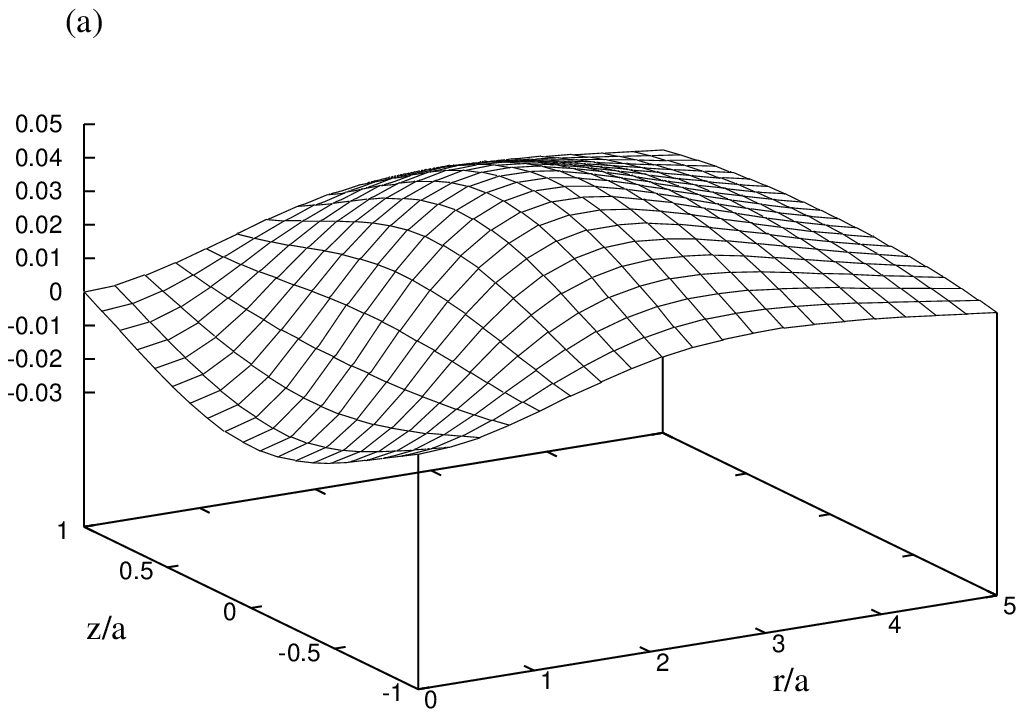} \ &
\ \epsfig{width=2.75in,file=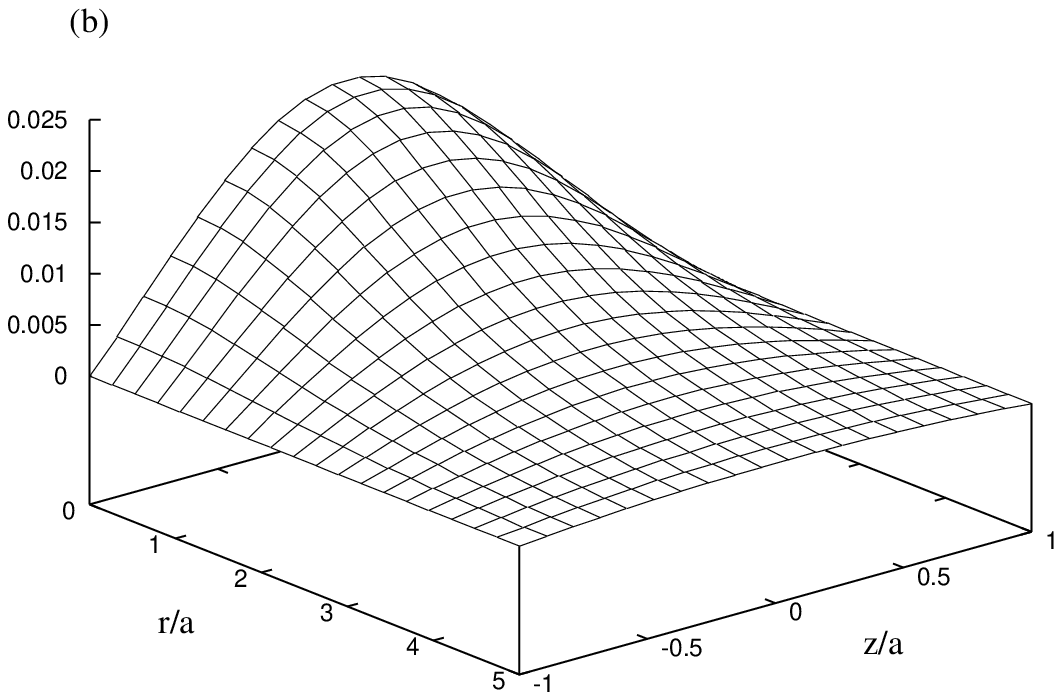} \ 
\end{array}$$
\caption{For a thick disk obtained from the Zipoy-Voorhees solution with
$\tilde m = 3$, $\tilde b = 3.5$ and $\tilde k = 1$ we plot (a) the azimuthal
stress ${\tilde p}_\varphi$ and (b) the vertical pressure ${\tilde
p}_z$.}\label{fig:pres2}
\end{figure*}

For any other value of $r$ and $z$ is not easy to obtain constraints over the
parameters $\tilde k$, $\tilde b$ and $\tilde m$ in order to have $\epsilon >
0$. The analysis is better done graphically. By considering different values of
$\tilde k$ and $\tilde b$ that fulfill the condition (\ref{eq:cond1}) we find
that $\epsilon > 0$ everywhere in the disks only if we take for $\tilde m$ a
value less than a tenth of the upper limit provided by the condition
(\ref{eq:cond2}). As an example, in Fig. \ref{fig:dens2} we plot the 
dimensionless densities $\tilde\rho$ and $\tilde\epsilon$ for a disk with
$\tilde m = 3$, $\tilde b = 3.5$ and $\tilde k = 1$. We can see that, as in the
Chazy-Curzon disk, the density have a maximum at the center of the $z = 0$
plane and then it decreases monotonously as $r$ increase 
and also for $z \to \pm a$. Also $\rho$ and $\epsilon$ have similar magnitudes.

The behavior of the stresses is also similar to the presented in the
Chazy-Curzon disk. Again, is better to do a graphical presentation. In Fig.
\ref{fig:pres2} we plot the  dimensionless azimuthal stress ${\tilde
p}_\varphi$ and the vertical pressure ${\tilde p}_z$ for the disk with $\tilde
m = 3$, $\tilde b = 3.5$ and $\tilde k = 1$. Again we have that $p_r = -
p_z$. As with the Chazy-Curzon disk, the azimuthal stress is negative at the
central region of the disk, then increase to have a positive maximum at the $z
= 0$ plane, for a value of $r \approx 2.5 a$, and finally decrease
monotonously. The behavior of the vertical pressure is like of the densities,
with a maximum at the center of the $z = 0$ plane and then monotonously
decreasing for increasing $r$. Also, $p_\varphi = 0$ for $z = \pm a$. From Fig.
\ref{fig:dens2} and \ref{fig:pres2} we have
\begin{eqnarray*}
\left| \frac{p_\varphi}{\epsilon} \right| &<& 0.2, \\
	\\
\left| \frac{p_r}{\epsilon} \right| = \left| \frac{p_z}{\epsilon} \right| &<&
0.1 ,
\end{eqnarray*}
and so the disks are also in agreement with the dominant energy condition. Thin
disks based in the Zipoy-Voorhees metric were considered in Ref. \cite{BLK}.

\section{Thick Disks from de Schwarzschild metric in isotropic coordinates}

For a static spherically symmetric spacetime the metric in isotropic spherical
coordinates $(t,R,\theta,\varphi)$ can be cast as
\begin{equation}
ds^2 = - e^{2\Phi} dt^2 + e^{2\Lambda} [ dR^2 + R^2 (d\theta^2 + \sin^2 \theta
d\varphi^2 ) ] , \label{eq:mets}
\end{equation} 
where $\Phi$ and $\Lambda$ are functions of $R$ only. In isotropic cylindrical
coordinates $(t,\varphi,r,z)$ the metric (\ref{eq:mets}) takes the form
\begin{equation}
ds^2 = - e^{2\Phi} dt^2 + e^{2\Lambda} [ r^2 d\varphi^2 + dr^2 + dz^2 ] ,
\label{eq:meti}
\end{equation} 
where now $\Phi$ and $\Lambda$ depends on $r$ and $z$.

We will now apply the ``displace, cut, fill and reflect'' method to
the Schwarzschild
solution in isotropic coordinates,
\begin{subequations}\begin{eqnarray}
\Phi & = & \ln \left[ \frac{2{\rm R} - m}{2{\rm R} + m} \right] , \\
& & 	\nonumber	\\
\Lambda & = & \ln \left[ 1 + \frac{m}{2{\rm R}} \right]^2 ,
\end{eqnarray}\label{eq:miso}\end{subequations}
 with $m$ a positive constant and ${\rm R}^2 = r^2 + z^2$.  Now we put
${\rm R}^2 = r^2 + (h+b)^2 $ where $b$ is a  positive constant and $h(z)$ is
given by (\ref{eq:funh}).

From  the Einstein equations (\ref{eq:eqm}) and the orthonormal tetrad 
$\{V^a, X^a, Y^a, Z^a\},$ where
\begin{subequations}\begin{eqnarray}
V^a &=& e^{- \Phi} \ ( 1, 0, 0, 0 ) \ ,	\\
	&	&	\nonumber	\\
X^a &=& \frac{e^\Phi}{r} \ \ ( 0, 1, 0, 0 ) \ ,	\\
	&	&	\nonumber	\\
Y^a &=& e^{- \Lambda} ( 0, 0, 1, 0 ) \ ,	\\
	&	&	\nonumber	\\
Z^a &=& e^{- \Lambda} ( 0, 0, 0, 1 ) \ ,
\end{eqnarray}\label{eq:tetrad2}\end{subequations}
we find that the energy-momentum tensor of the disk can be written as
\begin{equation}
T_{ab} = \epsilon V_a V_b + p_\varphi X_a X_b + p_r Y_a Y_b + p_z Z_a Z_b \ .
\label{eq:cans}
\end{equation}
Here $\epsilon = - T_t^t$ is the energy density, $p_\varphi =
T_\varphi^\varphi$ is the azimuthal stress, that is equal to the radial stress
$p_r = T_r^r$, and $p_z = T_z^z$ is the vertical stress. The effective
Newtonian density is given by $\rho = \epsilon + 2 p_r + p_z = \epsilon + 2
p_\varphi + p_z$.

From (\ref{eq:miso}) we obtain,  using the dimensionless variables
previously defined,
\begin{widetext}
\begin{subequations}\begin{eqnarray}
\tilde\rho &=& \frac{3 \tilde m \tilde{\rm R} \left[ 2 (3{\tilde z}^2 +
2(\tilde b - 1)) \tilde{\rm R}^2 + 3 ({\tilde z}^2 + 2\tilde b)^2 (1 - {\tilde
z}^2)\right]}{(2\tilde{\rm R} - \tilde m)(2\tilde{\rm R} + \tilde m)^5} ,
\label{eq:denis} \\
& & 	\nonumber	\\
\tilde\epsilon &=& \frac{3 \tilde m \left[2 (3{\tilde z}^2 + 2(\tilde b - 1))
\tilde{\rm R}^2 + 3 ({\tilde z}^2 + 2\tilde b)^2 (1 - {\tilde
z}^2)\right]}{2(2\tilde{\rm R} + \tilde m)^5} , \label{eq:eneis} \\
& & 	\nonumber	\\
{\tilde p}_\varphi &=& \frac{16 {\tilde m}^2 \left[(3{\tilde z}^2 + 2(\tilde b
- 1)) \tilde {\rm R}^2 + ({\tilde z}^2 + 2\tilde b)^2 (1 - {\tilde
z}^2)\right]}{(2\tilde{\rm R} - \tilde m)(2\tilde {\rm R} + \tilde m)^5} ,
\label{eq:p1is} \\
& & 	\nonumber	\\
{\tilde p}_r &=& \frac{16 {\tilde m}^2 \left[(3{\tilde z}^2 + 2(\tilde b - 1))
\tilde{\rm R}^2 + ({\tilde z}^2 + 2\tilde b)^2 (1 - {\tilde
z}^2)\right]}{(2\tilde{\rm R} - \tilde m)(2\tilde{\rm R} + \tilde m)^5} ,
\label{eq:p2is} \\
& & 	\nonumber	\\
{\tilde p}_z &=& \frac{16 {\tilde m}^2 ({\tilde z}^2 + 2\tilde b)^2 (1 -
{\tilde z}^2)}{(2\tilde{\rm R} - \tilde m)(2\tilde{\rm R} + \tilde m)^5} ,
\label{eq:p3is}
\end{eqnarray}\end{subequations}
\end{widetext}

From (\ref{eq:eneis}) and $\tilde b > 1$ we have  $\epsilon > 0$. On the other
hand, from (\ref{eq:denis}) and  $2 \tilde{\rm R} > \tilde m$ we have $\rho >
0$. Since  $\tilde{\rm R} \geq \tilde b$, this last condition is equivalent to
$2 \tilde b > \tilde m$. Therefore,  when $\tilde b \geq 1$ and $0 < \tilde m <
2 \tilde b$ we will have disks in agreement with the weak and strong energy
conditions. Also this values of $\tilde m$ assure the nonsingular behavior of
$\rho$, $p_\varphi$, $p_r$ and $p_z$. We also have $p_\varphi = p_r > 0$ and
$p_z > 0$. The vertical and horizontal stress are then pressures.

\begin{figure*}
$$\begin{array}{cc}
\ \epsfig{width=2.75in,file=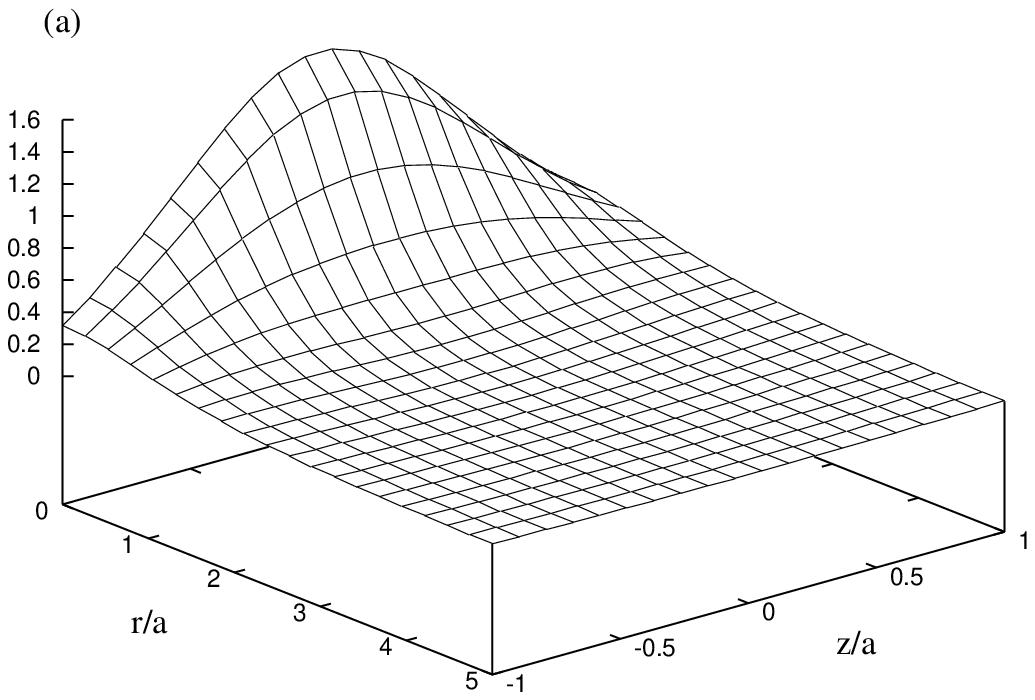} \ &
\ \epsfig{width=2.75in,file=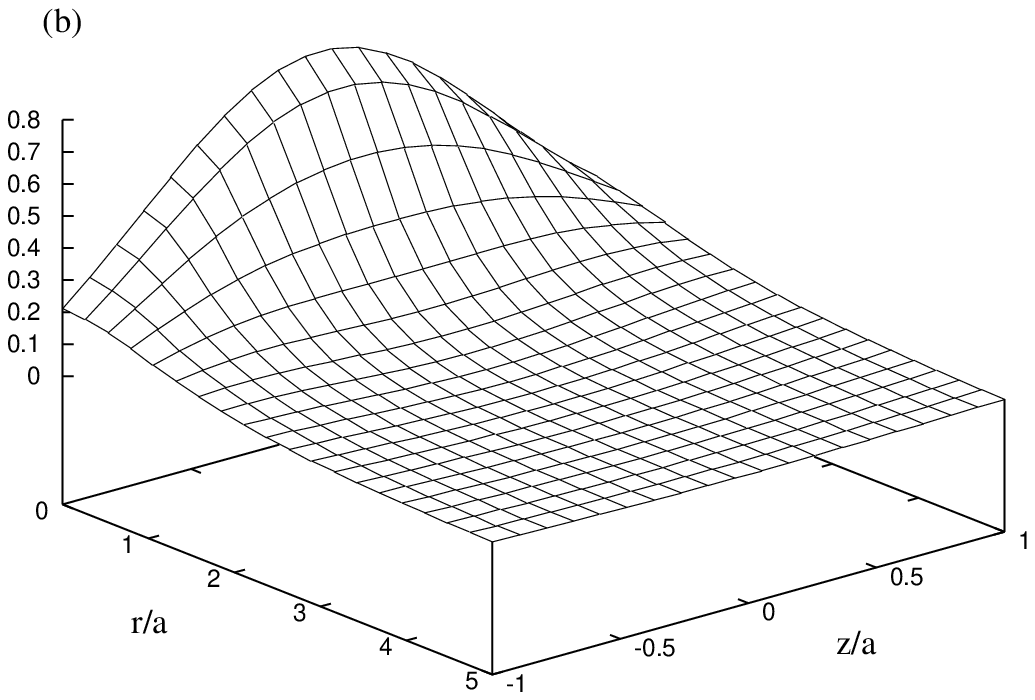} \ 
\end{array}$$
\caption{For a thick disk obtained from the Schwarzschild isotropic solution
with $\tilde m = \tilde b = 1$, we plot (a) the effective Newtonian density
$\tilde\rho$ and (b) the energy density $\tilde\epsilon$.}\label{fig:dens3}
\end{figure*}

\begin{figure*}
$$\begin{array}{cc}
\ \epsfig{width=2.75in,file=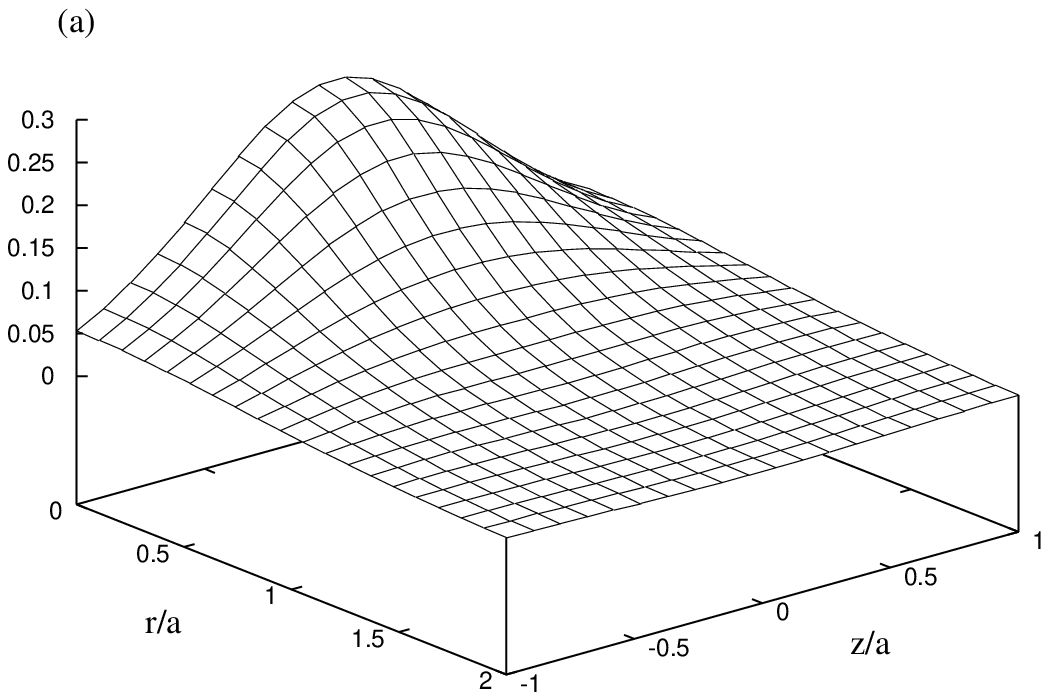} \ &
\ \epsfig{width=2.75in,file=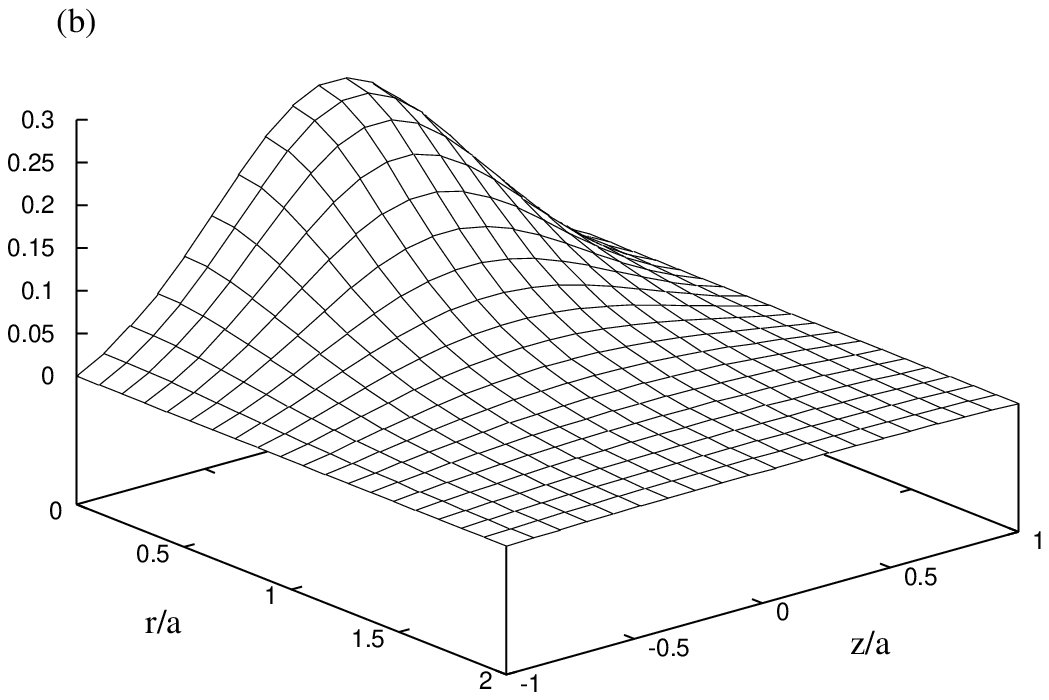} \ 
\end{array}$$
\caption{For a thick disk obtained from the Schwarzschild isotropic solution
with $\tilde m = \tilde b = 1$, we plot (a) the horizontal pressure ${\tilde
p}_\varphi = {\tilde p}_r$ and (b) the vertical pressure ${\tilde
p}_z$.}\label{fig:pres3}
\end{figure*}

As in the previous section, we perform a graphical analysis of the solution. In
Fig. \ref{fig:dens3} we plot $\tilde\rho$ and $\tilde\epsilon$ for a thick disk
obtained from the Schwarzchild isotropic solution with $\tilde m = \tilde b =
1$. The horizontal and vertical pressures are plotted in Fig \ref{fig:pres3}.
All the four quantities have a similar behavior, with a maximum at the center
of the $z = 0$ plane and then monotonously decreasing with increasing $r$ and
$z$. The relative magnitudes of the densities and pressures are such that $\rho
\geq \epsilon \geq p_\varphi = p_r \approx p_z$. We have
$$
\left| \frac{p_\varphi}{\epsilon} \right| = \left| \frac{p_r}{\epsilon} \right|
\approx \left| \frac{p_z}{\epsilon} \right| < 0.4 ,
$$
and so the disks are in agreement with all the energy conditions. Thin disks
based on the Schwarzschild solution in isotropic coordinates were studied in
\cite{DANLET},  whereas  thin disks based in the Schwarzschild metric in Weyl coordinates were studied in \cite{BLK}. For the disks in isotropic coordinates 
 we have matter with radial pressure equal to the azimuthal  pressure (isotropic matter) an for the disks in Weyl coordinates we have zero radial pressure.

\section{Discussion}

We presented a method to obtain exact general relativistic thick disks as a
generalization of the ``displace, cut and reflect'' method commonly used to
obtain Newtonian and relativistic thin disks. The generalization was be done by
means of the transformation $z \to h (z) + b$, where $h (z)$ is an even
function of $z$ and $b$ is a positive constant. The function $h(z)$ must be
selected in such a way that the metric tensor and and its first derivatives
will be continuous across the plane $z = 0$. 

All the cases considered leads to thick disks with similar behavior of the
energy and Newtonian effective densities: a maximum at the center of the
central plane of the disks, the $z = 0$ plane, and then monotonously decreasing
for increasing $r$ and $z$.

We found that, when the method is applied to vacuum Weyl spacetimes, the thick
disks presents radial tension and vertical pressure. The azimuthal stress is
negative at the central region of the disks, then have a positive maximum and
finally decrease monotonously for large values of $r$ and $z \to \pm a$, 
where $2a$ is the thickness of the disk. The disks obtained are in full
agreement with all the energy conditions.

On the other hand, when the method is applied to the Schwarzschild isotropic
metric all the stresses are pressures and have a behavior like the densities.
The disks obtained are also in full agreement with all the energy conditions.

 We plan  to extend the models of thick disk presented along this lines
 by considering more elaborate functions $h(z)$ and by the  incorporation of
 new properties like rotation, either electric or magnetic fields 
or both. Also we believe that the study of stability in this disks can produce some non trivial results.

\begin{acknowledgments}

We want to thank CNPq, FAPESP and COLCIENCIAS for financial support. Also G. A.
G. is grateful for the warm hospitality of the DMA-IMECC-UNICAMP where this
work was performed.

\end{acknowledgments}

\end{document}